\documentclass[11pt,preprint]{aastex}
\def\gtsima{$\; \buildrel > \over \sim \;$}
\def\gsim{\lower.5ex\hbox{\gtsima}}

\newdimen\minuswidth    
\setbox0=\hbox{$-$}
\minuswidth=\wd0
\catcode`@=\active
\def@{\kern\minuswidth}
\setbox0=\hbox{\rm0}
 
\shorttitle{The strong rotation of M5} 
\shortauthors{Lanzoni et al.}
 
\begin{document} 
\title{The strong rotation of M5 (NGC 5904) as seen from the MIKiS
  Survey of Galactic Globular Clusters\footnote{Based on FLAMES and
    KMOS observations performed at the European Southern Observatory
    as part of the Large Programme 193.D-0232 (PI: Ferraro).}}

\author{B. Lanzoni\altaffilmark{2,3}, F. R. Ferraro\altaffilmark{2,3},
  A. Mucciarelli\altaffilmark{2,3}, C. Pallanca\altaffilmark{2,3},
  E. Lapenna\altaffilmark{2,3}, L. Origlia\altaffilmark{3},
  E. Dalessandro\altaffilmark{3}, E. Valenti\altaffilmark{4},
  M. Bellazzini\altaffilmark{3}, M. A. Tiongco\altaffilmark{5},
  A. L. Varri\altaffilmark{6}, E. Vesperini\altaffilmark{5} and
  G. Beccari\altaffilmark{4}}

\affil{\altaffilmark{2} Dipartimento di Fisica e Astronomia,
  Universit\`a degli Studi di Bologna, via Gobetti 93/2, I$-$40129
  Bologna, Italy}
\affil{\altaffilmark{3}INAF-Osservatorio di Astrofisica e Scienza
  dello Spazio di Bologna, Via Gobetti 93/3, 40129, Bologna, Italy}
\affil{\altaffilmark{4}European Southern Observatory,
  Karl-Schwarzschild-Strasse 2, 85748 Garching bei M\"{u}nchen,
  Germany}
\affil{\altaffilmark{5} Dept. of Astronomy, Indiana University,
  Bloomington, IN, 47401, USA}
\affil{\altaffilmark{6} Institute for Astronomy, University of
  Edinburgh, Royal Observatory, Blackford Hill, Edinburgh EH9 3HJ, UK}
\date{25 April 2018; submitted to the ApJ}

\begin{abstract}
In the context of the ESO-VLT Multi-Instrument Kinematic Survey
(MIKiS) of Galactic globular clusters, we present the line-of-sight
rotation curve and velocity dispersion profile of M5 (NGC 5904), as
determined from the radial velocity of more than 800 individual stars
observed out to $700\arcsec$ ($\sim 5$ half-mass radii) from the
center. We find one of the cleanest and most coherent rotation
patterns ever observed for globular clusters, with a very stable
rotation axis (having constant position angle of $145\arcdeg$ at all
surveyed radii) and a well-defined rotation curve. The density
distribution turns out to be flattened in the direction perpendicular
to the rotation axis, with a maximum ellipticity of 0.15.  The
rotation velocity peak ($\sim 3$ km s$^{-1}$ in projection) is
observed at $\sim 0.6$ half-mass radii, and its ratio with respect to
the central velocity dispersion ($\sim 0.3$-0.4 at 4 projected
half-mass radii) indicates that ordered motions play a significant
dynamical role. This result strengthens the growing empirical evidence
of the kinematic complexity of Galactic globular clusters and
motivates the need of fundamental investigations of the role of
angular momentum in collisional stellar dynamics.
\end{abstract}
 

\section{INTRODUCTION}
\label{sec_intro}
Galactic globular clusters (GGCs) are ideal laboratories where the
large variety of phenomena due to collisional stellar dynamics can be
observationally studied.  Traditionally, they have been assumed to be
quasi-relaxed non-rotating stellar systems, characterized by spherical
symmetry and orbital isotropy.  Hence, spherical, isotropic and
non-rotating models, with a truncated distribution function close to
Maxwellian \citep{king66}, have been routinely used to fit the
observed surface brightness profiles and estimate the main structural
parameters and total mass \citep[e.g.][]{pryor+93, harris96,
  McLvdM05}.  However, recent N-body simulations indicate that GCs do
not attain complete energy equipartition (\citealp{trenti+13}; see
also \citealp{bianchini+16}) and they may show differential rotation
and complex behaviors of pressure anisotropy, depending on the degree
of dynamical evolution suffered and the effect of an external tidal
field \citep[e.g.][]{vesperini+14}.

Also from the observational point of view, increasing evidence is
demonstrating that these models are largely over-simplified. Indeed,
deviations from the sharply truncated King phase space distribution
(e.g., see the cases of NGC 1851, as studied by \citealp{olszewski+09,
  marino+14}, NGC 5694 by \citealp{correnti+11, bellazzini+15}, and
several others, as discussed, e.g., by \citealp{carballo-bello+18}),
spherical symmetry \citep[e.g.][]{chen+10} and pressure isotropy
\citep[e.g.][]{vandeven+06, bellini+14, bellini+17, watkins+15} are
found in a growing number of GGCs.  Also the observational evidence of
systemic rotation is increasing (e.g., \citealt{anderson+03, lane+09,
  lane+10, bellazzini+12, bianchini+13, fabricius+14, kacharov+14,
  lardo+15, kimmig+15, bellini+17, boberg+17, cordero+17, kamann+18,
  ferraro+18a}), possibly suggesting that, when properly surveyed, the
majority of GCs rotate at some level. In particular,
\citet{ferraro+18a} investigated the intermediate/external region of
11 clusters, finding evidence of rotation within a few half-mass radii
from the center in 9 systems.  \citet{kamann+18} surveyed the central
regions of 25 GGCs, detecting signals of rotation in $60\%$ of their
sample.  On the other hand, recent N-body simulations
\citep{tiongco+17} describing the long-term evolution of GC rotational
properties suggest that the detection of (even modest) signals is
crucial, since these could be the relic of significant internal
rotation set at the epoch of the cluster's formation.
 
As part of the ESO-VLT Multi-Instrument Kinematic Survey of Galactic
Globular Clusters (hereafter the MIKiS Survey;
\citealp{ferraro+18a}),\footnote{The MIKiS Survey was specifically
  designed to provide velocity dispersion and rotation profiles from
  the radial velocity of hundred individual stars for a representative
  sample of GGCs, by exploiting the spectroscopic capabilities of
  different instruments (SINFONI+KMOS+FLAMES) available at the ESO
  Very Large Telescope (VLT).}  here we present the line-of-sight
internal kinematics of M5 (NGC 5904) obtained from the combination of
FLAMES and KMOS data.  With a total sample of more than 800 stars
extending out to $\sim 5$ half-mass radii, the dataset presented here
allowed us to construct the most detailed rotation curve and velocity
dispersion profile so far for the intermediate/outer regions of the
system, clearly showing the presence of a coherent systemic rotation
pattern. The paper is organized as follows. In Section \ref{sec_obs}
we describe the observational dataset and the data reduction
procedures adopted for the analysis.  The determination of the radial
velocity (RV) from the acquired individual star spectra is discussed
in Section \ref{sec_RV}.  Section \ref{sec_resu} is devoted to present
the obtained results: the systemic radial velocity of the system, its
rotation curve and velocity dispersion profile, and the projected
density map determined from resolved star photometry, from which we
estimated the cluster ellipticity. The results are then discussed in
Section \ref{sec_discuss}.

\section{Observations and data reduction}
\label{sec_obs}
The observational strategy and the data reduction procedure adopted in
the MIKiS Survey are described in \citet{ferraro+18a}. Here we
schematically remind just the main points.

We used FLAMES (\citealt{pasquini+00}) in the GIRAFFE/MEDUSA combined
mode (consisting of 132 deployable fibers which can be allocated
within a $25\arcmin$-diameter field of view), adopting the HR21
grating setup, with a resolving power R$\sim 16200$ and a spectral
coverage from 8484 \AA{} to 9001 \AA. This grating samples the
prominent Ca II triplet lines, which are excellent features to measure
RVs.  The target stars have been selected from Hubble Space Telescope
ACS/WFC data acquired in the F606W and F814W bands
\citep{sarajedini+07} and a complementary wide-field catalog in $B$
and $V$ obtained from ESO-WFI observations, as described in
\citet{lan07_m5}.  They are located along the red giant, asymptotic
giant and horizontal branches of the cluster, at magnitudes brighter
than $V=17.0$ ($V_{\rm ground}$ in the ACS catalog). To prevent the
contamination of the target spectra from close sources, only stars
with no bright neighbors ($V_{\rm neighbor}< V_{\rm star}+1.0$) within
$2\arcsec$ have been selected.  We secured five different pointings
with total integration times ranging from 900 s to 1800 s, according
to the magnitude of the targets.  In each exposure, typically 15-20
spectra of the sky were acquired; these have been averaged to obtain a
master sky spectrum, which was then subtracted from the spectrum of
each target.  For homogeneization purposes, we re-observed $\sim 30$
stars in common with the pre-existing FLAMES datasets of M5 that we
retrieved from the ESO archive (see Table \ref{tab_datasets}).  The
data reduction of both the MIKiS Survey exposures and the archive
spectra was performed by using the FLAMES-GIRAFFE
pipeline,\footnote{http://www.eso.org/sci/software/pipelines/} which
includes bias-subtraction, flat-field correction, wavelength
calibration with a standard Th-Ar lamp, re-sampling at a constant
pixel-size and extraction of one-dimensional spectra.

KMOS \citep{sharples+10} is a spectrograph equipped with 24 deployable
IF units, each of $2.8\arcsec\times2.8\arcsec$ on the sky, that can be
allocated within a $7.2\arcmin$ diameter field of view.  We have used
the YJ grating covering the 1.00-1.35 $\mu$m spectral range at a
resolution R$\approx$3400.  This setup is especially effective in
simultaneously measuring a number of reference telluric lines in the
spectra of giant stars, for an accurate calibration of the RV, despite
the relatively low spectral resolution.  The selected targets are red
and asymptotic giant branch stars with $J<14$ ($V<16.8$), located
within $\sim 145\arcsec$ from the cluster center.  For a proper
homogeneization of the RV measures, $\sim 30$ targets have been
selected in common with the FLAMES dataset.  We secured ten pointings
with total integration times ranging from 30 s to 100 s, depending on
the target magnitudes. The typical SNR of the observed spectra is
$\gtrsim$ 50.  We used the ``nod to sky'' KMOS observing mode and
nodded the telescope to an off-set sky field at $\approx 6\arcmin$
North of the cluster center, for a proper background subtraction.  The
raw data have been reduced using the KMOS pipeline$^3$ which performs
background subtraction, flat-field correction and wavelength
calibration of the 2D spectra.  The 1D spectrum from the brightest
spaxel of each target star centroid was then extracted manually.

\section{Radial velocity measurements}
\label{sec_RV}
RVs were obtained as described in \citet{ferraro+18a}.  In short, we
followed the procedure discussed in \citet{tonry+79},
cross-correlating the observed spectra (corrected for heliocentric
velocity) with a template of known velocity.  As templates we used
synthetic spectra computed with the SYNTHE code \citep[see
  e.g.][]{sbordone+04}, adopting the cluster metallicity and
appropriate atmospheric parameters according to the evolutionary stage
of the targets. The typical uncertainties in the RVs derived from
FLAMES spectra are of the order of 0.1-0.5 km s$^{-1}$.  Uncertainties
in the RVs derived from KMOS spectra have been estimated through
Montecarlo simulations, using cross-correlation against synthetic
spectra of appropriate metallicity, opportunely resampled at the KMOS
pixel-scale, and with Poissonian noise added.  We created 500 noisy
spectra for different SNR values in the range between about 30 and
100.  The RVs of these samples have been measured by using the
cross-correlation technique adopted for the observed KMOS spectra, and
the dispersion of the derived RVs has been assumed as the typical RV
uncertainty ($\epsilon_{\rm RV}$) for the corresponding SNR. The
derived relation between SNR and RV error is: $\ln(\epsilon_{\rm
  RV})=6.231 - 1.169 \, \ln({\rm SNR})$.  The stars in common were
used to report the KMOS and the archive measures to the MIKiS RVs
(determined from the HR21 grating).

If multiple exposures were available for the same star, we adopted the
RV obtained from the weighted mean of the highest resolution measures,
by using the individual errors as weights (hence, for all the stars in
common between the FLAMES and the KMOS data sets, we adopted the
FLAMES values).
 
\section{Results}
\label{sec_resu}
\subsection{Systemic velocity}
\label{sec_vsys}
\begin{figure}
\includegraphics[width=15cm]{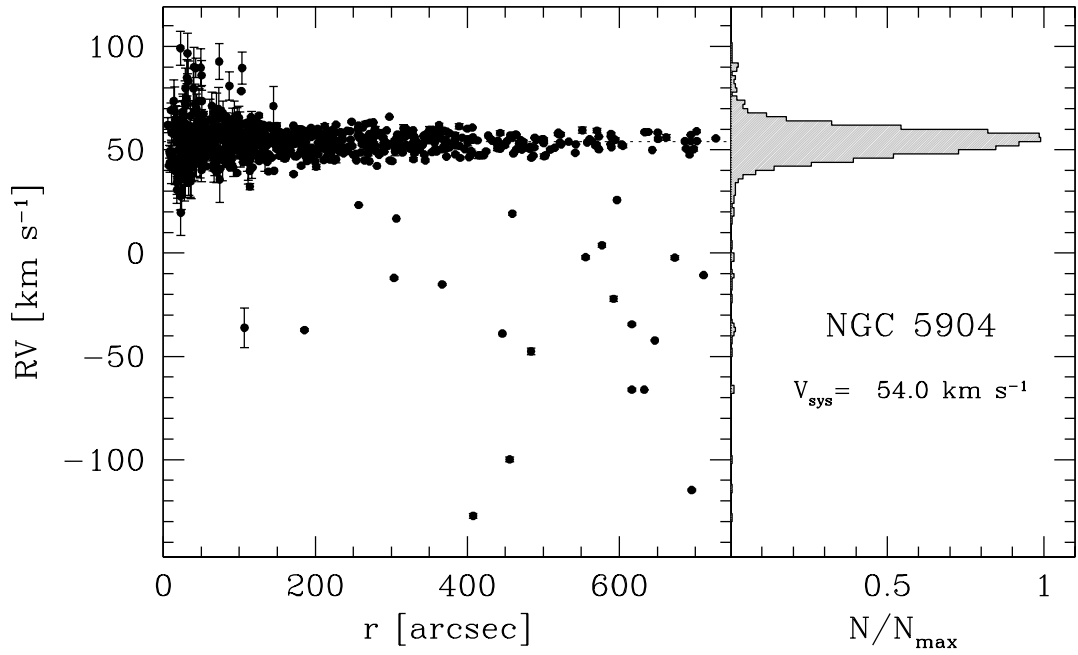}
\caption{{\it Left panel:} Observed radial velocities as a function of
  the distance from the cluster center obtained in this wok.  {\it
    Right panel:} histogram of the RV distribution, normalized to its
  peak value. The value of the derived systemic velocity of M5 is
  labelled.}
\label{fig_vrr}
\end{figure}

The final sample of RVs in the direction of M5 consists of 857
measures for individual sources distributed out to $727\arcsec$ from
the cluster center. Adopting the values quoted in \citet{miocchi+13},
this corresponds to $\sim 26$ core radii ($r_c=28\arcsec$) or 5
half-mass radii ($r_h=140\arcsec$). The innermost star is at
$r=6\arcsec$, but only a dozen of measures are available within
15-$20\arcsec$ from the center because of the stellar crowding
limitations.  The distribution of RVs as a function of the distance
from the center is plotted in Figure \ref{fig_vrr}.  The population of
cluster members is clearly distinguishable as a narrow and strongly
peaked distribution, while the Galactic field component is negligible
at all radii.  Assuming that the RV distribution is Gaussian, we used
a Maximum-Likelihood approach \citep[e.g.,][]{walker06} to estimate
cluster systemic velocity and its uncertainty.  For this purpose only
the 677 RVs measured from FLAMES spectra have been used, and obvious
outliers (as field stars) have been excluded from the analysis by
means of a $3\sigma$-clipping procedure. The resulting value of the
cluster systemic velocity is $V_{\rm sys} =54.0 \pm 0.2$ km s$^{-1}$,
in good agreement with previous determinations (see \citealt{harris96,
  kimmig+15}).  In the following, we will use $V_r$ to indicate RVs
referred to the cluster systemic velocity: $V_r\equiv {\rm RV} -V_{\rm
  sys}$.

\subsection{Systemic rotation}
\label{sec_vrot}
\begin{figure}
\includegraphics[width=15cm]{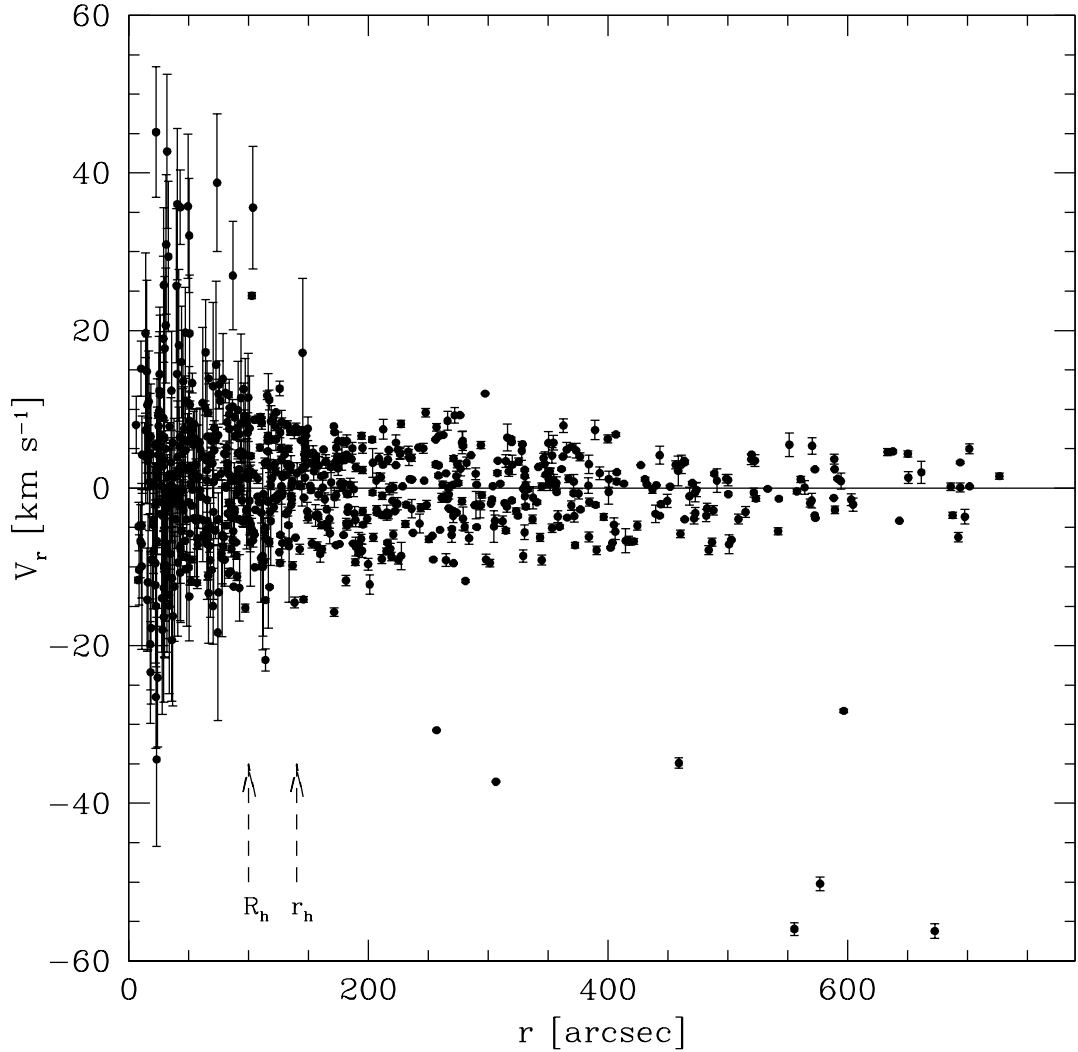}
\caption{{\it Left panel:} Zoomed view of the radial distribution of
  the measured velocities referred to $V_{\rm sys}$. The large scatter
  observable at large distances from the cluster center is a clear
  signature of systemic rotation. The two arrows indicate the
  projected half-light (or, equivalently, half-mass) radius and the
  three-dimensional half-mass radius of M5 ($R_h =100\arcsec$ and
  $r_h=140\arcsec$, respectively; from \citealp{miocchi+13}).}
\label{fig_vrr_zoom}
\end{figure}

\begin{figure}
\includegraphics[width=15cm]{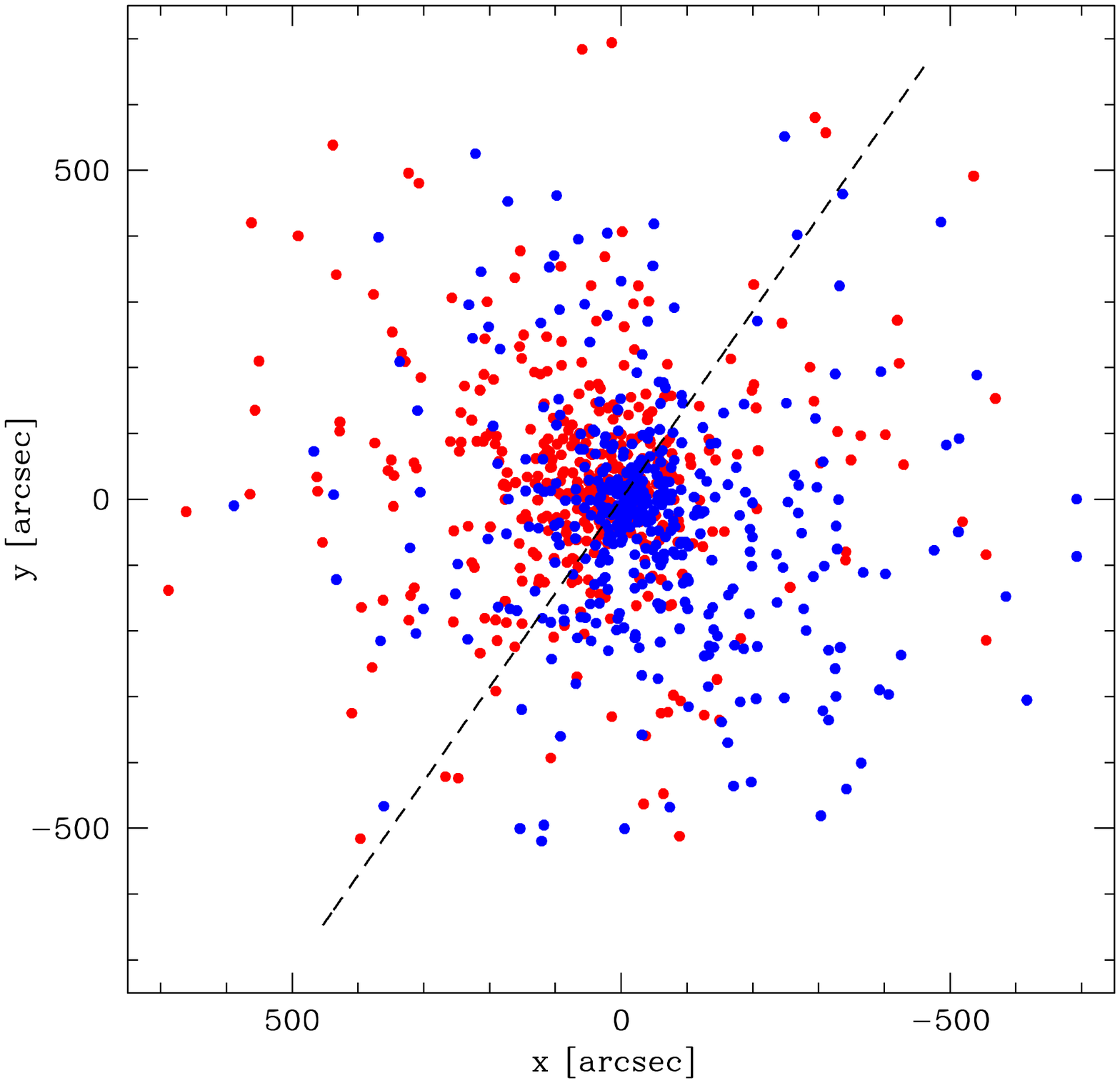}
\caption{Distribution of the observed sample on the plane of the sky,
  with $x= {\rm (RA-RA_0) \cos(Dec)}$ and $y={\rm Dec-Dec_0}$
  (RA${_0}$ and Dec${_0}$ being the coordinates of the cluster center,
  adopted from \citealp{miocchi+13}). North is up, East is on the
  left. The colors distinguish stars with radial velocity larger than
  $V_{\rm sys}$ (in red), from those with $V_r<0$ (in blue). The
  dashed line marks the position of the rotation axis, which has a
  position angle of $145\arcdeg$ from North (as measured
  anti-clockwise).  }
\label{fig_map}
\end{figure}

A zoomed view of $V_r$ as a function of the distance from the center
(Figure \ref{fig_vrr_zoom}) clearly shows that at the outermost
sampled radii (corresponding to 5 $r_h$) the distribution remains
broad about the cluster systemic velocity. This is not expected for an
isotropic pressure-supported system, where the velocity dispersion
formally decreases to zero in the outskirts. It can be explained,
instead, as an effect of systemic rotation.  Figure \ref{fig_map}
shows the distribution of the surveyed stars on the plane of the sky
(where $x$ and $y$ are the RA and Dec coordinates referred to those of
the cluster center, adopted from \citealp{miocchi+13}), with the red
and the blue colors indicating, respectively, positive and negative
values of $V_r$ (i.e., RVs larger and smaller than the systemic
velocity, respectively). As apparent from the figure, the evident
prevalence of stars with positive values of $V_r$ in the upper-left
portion of the map, and that of sources with $V_r<0$ in the
lower-right part of the diagram is clear-cut signature of systemic
rotation.

To investigate the rotation properties in this cluster we used the
same approach adopted in \citet{ferraro+18a} and described, e.g., in
\citet[][see also \citealp{lanzoni+13}]{bellazzini+12}.  The method
consists in splitting the RV dataset in two sub-samples with a line
passing through the cluster center, and determining the difference
between the mean velocity of the two groups ($\Delta V_{\rm
  mean}$). This is done by varying the position angle (PA) of the
splitting line from $0\arcdeg$ (North direction) to $180\arcdeg$
(South direction), by steps of $10\arcdeg$, and with $90\arcdeg$
direction corresponding to the East.  In the presence of rotation,
$\Delta V_{\rm mean}$ draws a coherent sinusoidal variation as a
function of PA, its maximum absolute value providing twice the
rotation amplitude ($A_{\rm rot}$) and the position angle of the
rotation axis (PA$_0$).  The rotation of the standard coordinate
system with respect to the cluster center ($x,y$) over the position
angle PA$_0$ provides the rotated coordinate system (XR,YR), with XR
set along the cluster major axis and YR aligned with the rotation
axis.  In a diagram showing $V_r$ as a function of the projected
distances from the rotation axis (XR) the stellar distribution shows
an asymmetry, with two diagonally opposite quadrants being more
populated than the remaining two. Moreover, the sub-samples of stars
on each side of the rotation axis (i.e., with positive and with
negative values of XR) have different cumulative $V_r$ distributions
and different mean velocities. To quantify the statistical
significance of such differences we used three estimators: the
probability that the RV distributions of the two sub-samples are
extracted from the same parent family is evaluated by means of a
Kolmogorov-Smirnov test, while the statistical significance of the
difference between the two sample means is estimated with both the
Student's t-test and a Maximum-Likelihood approach.

\begin{figure}
\includegraphics[width=15cm]{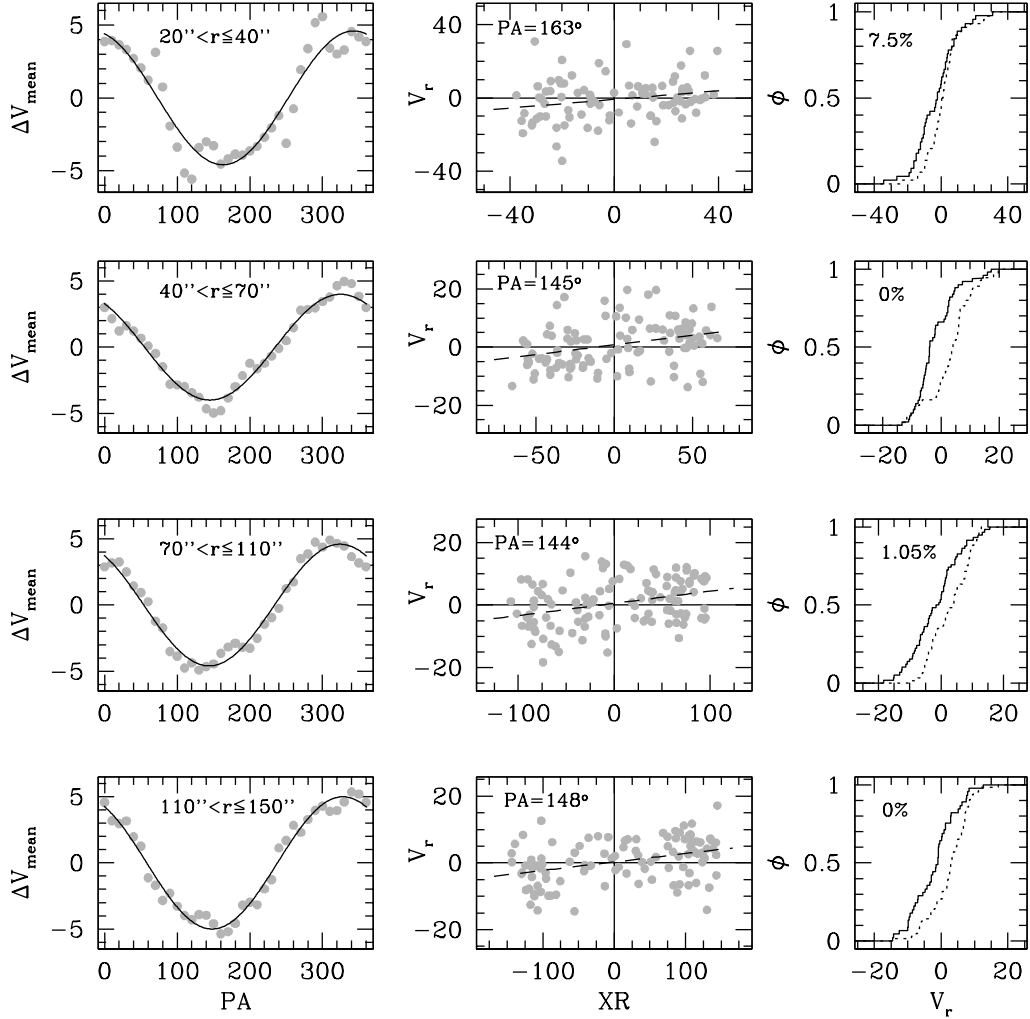}
\caption{Diagnostic diagrams of the rotation signature detected in the
  first four concentric annuli listed in Table \ref{tab_vrot_anelli}
  (see labels in the left-hand panels).  For each bin the {\it left
    panel} shows the difference between the mean RV on each side of a
  line passing through the center with a given position angle (PA), as
  a function of PA itself. The continuos line is the sine function
  that best fits the observed pattern.  The {\it central panel} shows
  the distribution of the radial velocities $V_r$ as a function of the
  projected distances from the rotation axis (XR) in arcseconds. The
  position angle of the rotation axis is labelled. The dashed line is
  the least square fit to the data.  The {\it right panel} shows the
  cumulative RV distributions for the sample of stars with XR$<0$
  (solid line) and for that with XR$>0$ (dotted line). The
  Kolmogorov-Smirnov probability that the two samples on each side of
  the rotation axis are drawn from the same parent distribution is
  labelled.}
\label{fig_vrot_anelli1}
\end{figure}

\begin{figure}
\includegraphics[width=15cm]{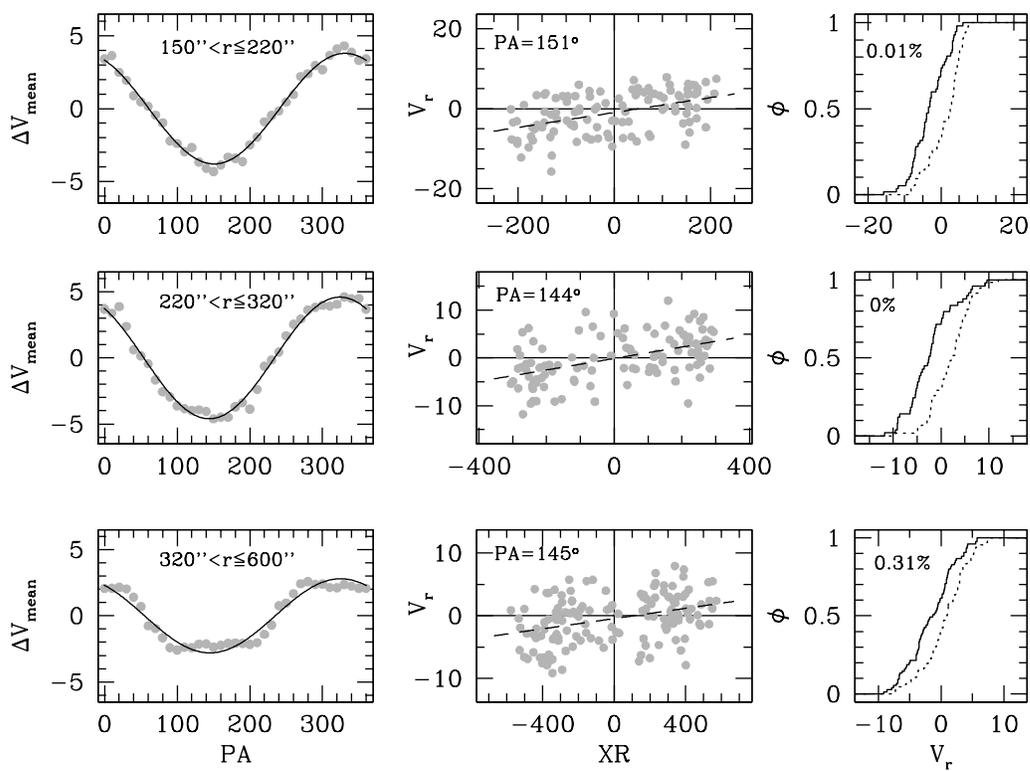}
\caption{As in Fig. \ref{fig_vrot_anelli1}, but for the three
  outermost considered annuli (see the labels in the left-hand
  panels).}
\label{fig_vrot_anelli2}
\end{figure}

We applied this procedure to our RV sample in a set of concentric
annuli around the cluster center, avoiding the innermost region
($r<20\arcsec$), where the statistic is poor, and the outermost region
($r>600\arcsec$), where the sampling is scant and non-symmetric.  The
results are listed in Table \ref{tab_vrot_anelli} and plotted in
Figures \ref{fig_vrot_anelli1} and \ref{fig_vrot_anelli2}. In all the
considered annuli, we find well-defined sinusoidal behaviors of
$\Delta V_{\rm mean}$ as a function of PA (left-hand panels in
Figs. \ref{fig_vrot_anelli1} and \ref{fig_vrot_anelli2}), asymmetric
distributions of $V_r$ as a function of the projected distance from
the rotation axis XR (central panels), and well-separated cumulative
$V_r$ distributions for the two samples on either side of the rotation
axis (right-hand panels). The reliability of these systemic rotation
signatures is also confirmed by the values of the Kolmogorov-Smirnov
and t-Student probabilities and by the significance level of different
sample means obtained from the Maximum-Likelihood approach (see the
thee last columns in Table \ref{tab_vrot_anelli}). Furthermore, as
also shown in Figure \ref{fig_pa}, the position angle of the rotation
axis (PA$_0$) is essentially constant in all the investigated annuli,
as expected in the case of a coherent global rotation of the system.
To conservatively determine the best-fit position angle PA$_0$ of the
global rotation of M5 we considered only the radial range
($r>40\arcsec$) where statistically significant signatures are
detected.  We thus found PA$_0=145\arcdeg$.  Its location in the plane
of the sky ($x,y$) is shown as a dashed line in Figures \ref{fig_map}
and \ref{fig_pa}. By fixing PA$_0$ to this value and using all the
observed stars, we finally obtain the diagnostic plots shown in Figure
\ref{fig_vrot_glob_PA145} and the values listed in Table
\ref{tab_vrot_glob_PA145} for the global rotation signatures of M5.
This is one of the strongest and cleanest evidence of rotation found
to date in a GC.

\subsection{Ellipticity}
\label{sec_dens}
A rapidly rotating system also is expected to be flattened in the
direction perpendicular to the rotation axis
\citep{chandrasekhar69}. To investigate this issue we used the HST/ACS
and ESO-WFI catalogs discussed above and build the stellar density map
of the system. Only stars with $V<19$ ($\sim 0.5$ magnitudes below the
main sequence turn-off point) have been used to avoid incompleteness
effects. This allowed us to extend the analysis out to $\sim
200\arcsec$.  The resulting map is shown in Figure \ref{fig_dens},
where the black solid lines draw the isodensity contours, the white
lines correspond to their best-fit ellipses and the dashed straight
line marks the position of the rotation axis.

As apparent, the stellar density distribution has spherical symmetry
in the center and becomes increasingly flattened in the direction
perpendicular to the rotation axis for increasing radius. This trend
is qualitatively consistent with that predicted, for example, by the
models introduced by \citet{varri+12} and found in the observational
study of 47 Tucanae \citep{bianchini+13, bellini+17}. For the two
outermost ellipses shown in the figure (at $r\sim 80\arcsec$ and
$r\sim 120\arcsec$) we measure $1-b/a= 0.1$ and 0.15, where $a$ and
$b$ are the major and the minor axes, respectively.

\subsection{Rotation curve and velocity dispersion profile}
\label{sec_kin}
To determine the rotation curve of M5 we considered the rotated
coordinate system (XR,YR) and split the $V_r$ sample in five intervals
of XR on both sides of the rotation axis. We then used the
Maximum-Likelihood method described in \citet[][see also Martin et
  al. 2007; Sollima et al. 2009]{walker06} to determine the mean
velocity of all the stars belonging to each XR bin.  The errors have
been estimated following \citet{pryor+93}.  The resulting rotation
curve (Figure \ref{fig_rotcurve} and Table \ref{tab_kin}) clearly
shows the expected shape, with an increasing trend in the innermost
regions up to a maximum value, and a decreasing behavior outward.  The
analytic expression \citep{lyndenbell67} appropriate for cylindrical
rotation:
\begin{equation}
  V_{\rm rot} = \frac{2 A_{\rm peak}}{\rm XR_{peak}} ~~ \frac{\rm
    XR}{1 + ({\rm XR}/{\rm XR_{peak}})^2}
\label{eq_curve}  
\end{equation}
very well reproduces the observed rotation curve (see the red solid
line in Figure \ref{fig_rotcurve}), with a maximum amplitude of $\sim
3$ km s$^{-1}$ at $\sim 90\arcsec$ from the rotation axis.

By folding the two RV samples on either side of the rotation axis and
using the same five intervals of XR adopted for the rotation curve, we
obtained the projected velocity dispersion profile shown in the
left-hand panel of Figure \ref{fig_vdisp} and listed in the last two
columns of Table \ref{tab_kin}.  We emphasize that the velocity
dispersion profiles most commonly shown in the literature are
determined in circular annuli around the cluster center, rather than
in shells of projected radial distances from the rotation axis (XR),
as done in this figure.  However, in the presence of a clear global
rotation of the system, it is reasonable to assume cylindrical
symmetry and thus to show the kinematical properties in the rotated
coordinate system (XR,YR). Indeed, this allows a direct comparison
with the rotation velocity (Fig. \ref{fig_rotcurve}), which is
determined in the same projection.  This comparison clearly shows
that, in spite of a clean and relatively strong rotation, M5 is still
dominated by non-ordered motions at all distances from the rotation
axis: in fact, the velocity dispersion is larger than the rotation
velocity in all the considered bins.

The projected velocity dispersion profile of M5 obtained in circular
concentric shells is shown in the right-hand panel of Figure
\ref{fig_vdisp} (black circles) and listed in Table \ref{tab_vdisp}.
This has been determined after subtracting from the measured RV of
each star, the mean velocity of the XR shell to which the star
belongs.  For sake of illustration we also show the radial profile of
second velocity moment (grey circles), i.e., the dispersion of the RVs
measured within each circular bin, with no subtraction of the
rotational component.  Of course, the velocity dispersion is smaller
than the second velocity moment in every bin. However, the differences
are small and always within the errors, as expected in the case of a
pressure-supported system.  The comparison between the left-hand panel
and the right-hand panel (black circles) of Fig. \ref{fig_vdisp}
clearly shows that the central values of the velocity dispersion
obtained by using XR shells are smaller than those determined in
circular annuli.  This is due to the fact that, by construction, the
inner XR shells include stars that are spatially close to the rotation
axis, but orbit both in the cluster central regions (hence, with large
velocity dispersion) and at the cluster periphery (hence, with small
velocity dispersion).  The innermost circular annulus, instead, is
largely dominated by stars that are truly orbiting close to the
center, and the ``dilution'' effect due to physically distant stars is
much smaller.

Since our observations extend out $\sim 10\arcmin$ away from the
center, the projection of the cluster space motion along the
line-of-sight could produce a non-negligible amount of apparent
rotation.  To estimate the contribution of such perspective rotation
to the true rotational velocity of M5 we followed the procedure
described in \citet{vandeven+06}, adopting the values quoted in
\citet{narloch+17} for the systemic proper motion of M5. We found a
mild variation of the position angle of the rotation axis
(PA$_0=141\arcdeg$, instead of $145\arcdeg$) and values of $V_{\rm
  rot}$ and $\sigma_P$ in very good agreement (well within the errors)
with those quoted in Table \ref{tab_kin}.  These results and the fact
that updated values of the cluster proper motion will become available
soon (thanks to the upcoming Gaia second data release), we decide not
to apply perspective rotation corrections to our determinations.

\section{Discussion}
\label{sec_discuss}
As part of the ESO-VLT MIKiS Survey \citep{ferraro+18a}, we presented
solid and unambiguous evidence of strong global rotation between $\sim
0.5 r_h$ and $5 r_h$ in the Galactic globular cluster M5. Signatures
of systemic rotation in this system, both in the outskirts and in the
central regions, were already presented in previous
works. \citet{bellazzini+12} found a rotation signal, with an
amplitude of 2.6 km s$^{-1}$ and a position angle of $157\arcdeg$,
from the analysis of 136 individual star spectra at $\sim 60\arcsec
<r<600\arcsec$. From a sample of 128 stars distributed between $\sim
70\arcsec$ and $\sim 1400\arcsec$, \citet{kimmig+15} report an
amplitude of 2.1 km s$^{-1}$. \citet{fabricius+14} performed an
integrated-light spectroscopic study of the innermost $\sim
60\arcsec\times 60\arcsec$ of M5, finding a central velocity gradient
of 2.1 km s$^{-1}$ and a position angle of the rotation axis of
$148.5\arcdeg$ (once reported in the coordinate system adopted
here). Very recently, \citet{kamann+18} analyzed a large number of
individual star spectra acquired at $r<60\arcsec$ with the
integral-field spectrograph ESO-MUSE, and found a velocity gradient of
2.2 km s$^{-1}$. They measured the rotation axis position angle in
different radial bins around the cluster center, finding PA$_0\sim
130\arcdeg-140\arcdeg$ (once reported in our system) at $r>10\arcsec$,
while the axis seems to be rotated by $90\arcdeg$ in the innermost
region. Although a detailed comparison of the rotation amplitude among
the various works is not straightforward (because of the different
radial regions sampled and/or the different parameters adopted to
quantify it), typical values of $\sim 2$ km s$^{-1}$ are found in all
the studies. A very good agreement is also found for what concerns the
position angle of the rotation axis. The only exception is the
perpendicular direction found by \citet{kamann+18} in the innermost
$10\arcsec$ of the cluster.  Higher spatial-resolution spectroscopy,
with the enhanced version of MUSE (WFM-AO, which operates under
super-seeing conditions down to FWHM$\sim 0.4\arcsec$), or with the
adaptive-optics corrected spectrograph ESO-SINFONI (see
\citealp{lanzoni+13}), will shed new light on this intriguing feature.

\begin{figure}
\includegraphics[width=15cm]{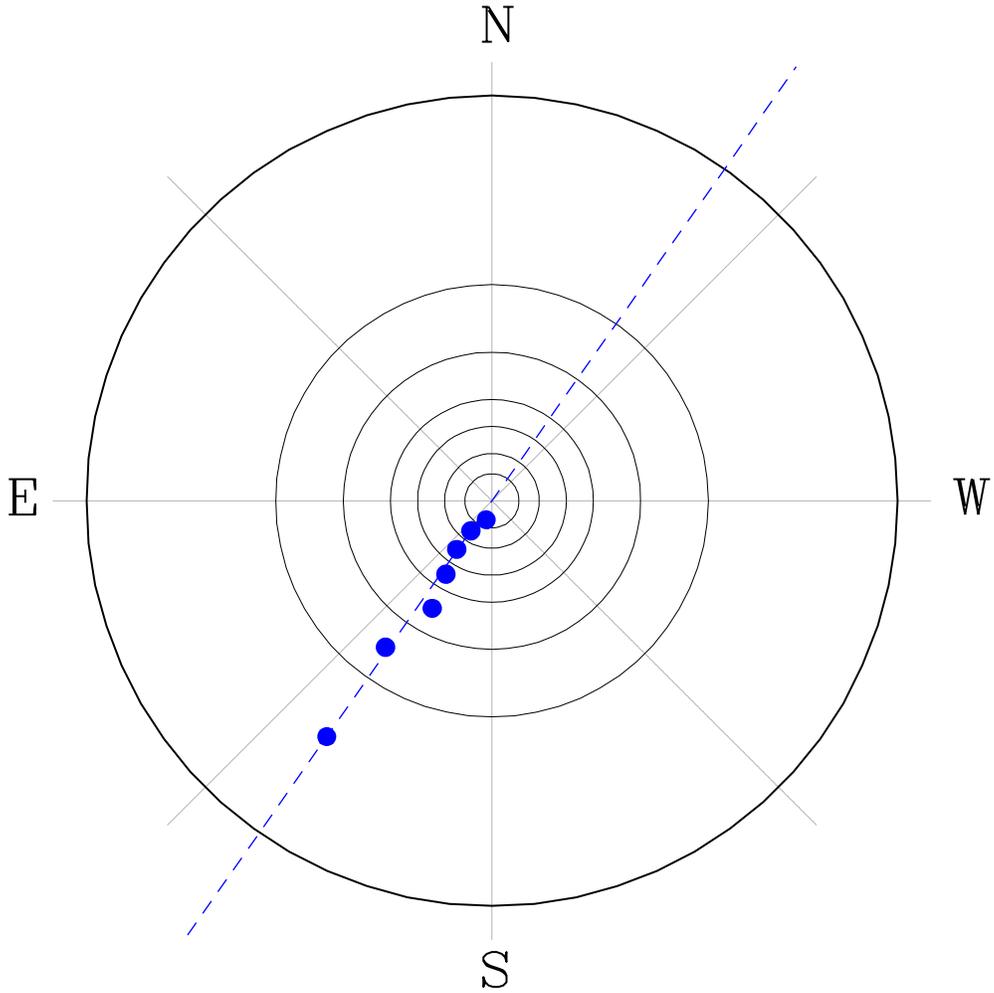}
\caption{Position angle of the rotation axis in the radial bins listed
  in Table \ref{tab_vrot_anelli}, used to search for rotation
  signatures (blue circles). As apparent, the value of PA is constant
  in all bins. The dashed line marks the direction of the adopted
  rotation axis of M5 (with position angle of $145\arcdeg$, as
  measured anti-clockwise from North to South).}
\label{fig_pa}
\end{figure}

\begin{figure}
\includegraphics[width=15cm]{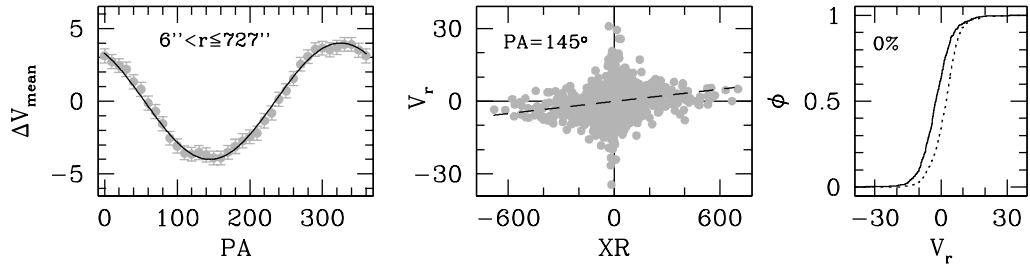}
\caption{Diagnostic diagrams of the global rotation of M5. The meaning
  of each panel is as in Figs. \ref{fig_vrot_anelli1} and
  \ref{fig_vrot_anelli2}, but here we plot all the observed stars
  (with cluster-centric distances $6\arcsec <r\le 727\arcsec$),
  assuming PA$_0=145\arcdeg$ as position angle of the rotation axis.}
\label{fig_vrot_glob_PA145}
\end{figure}

\begin{figure}
\includegraphics[width=15cm]{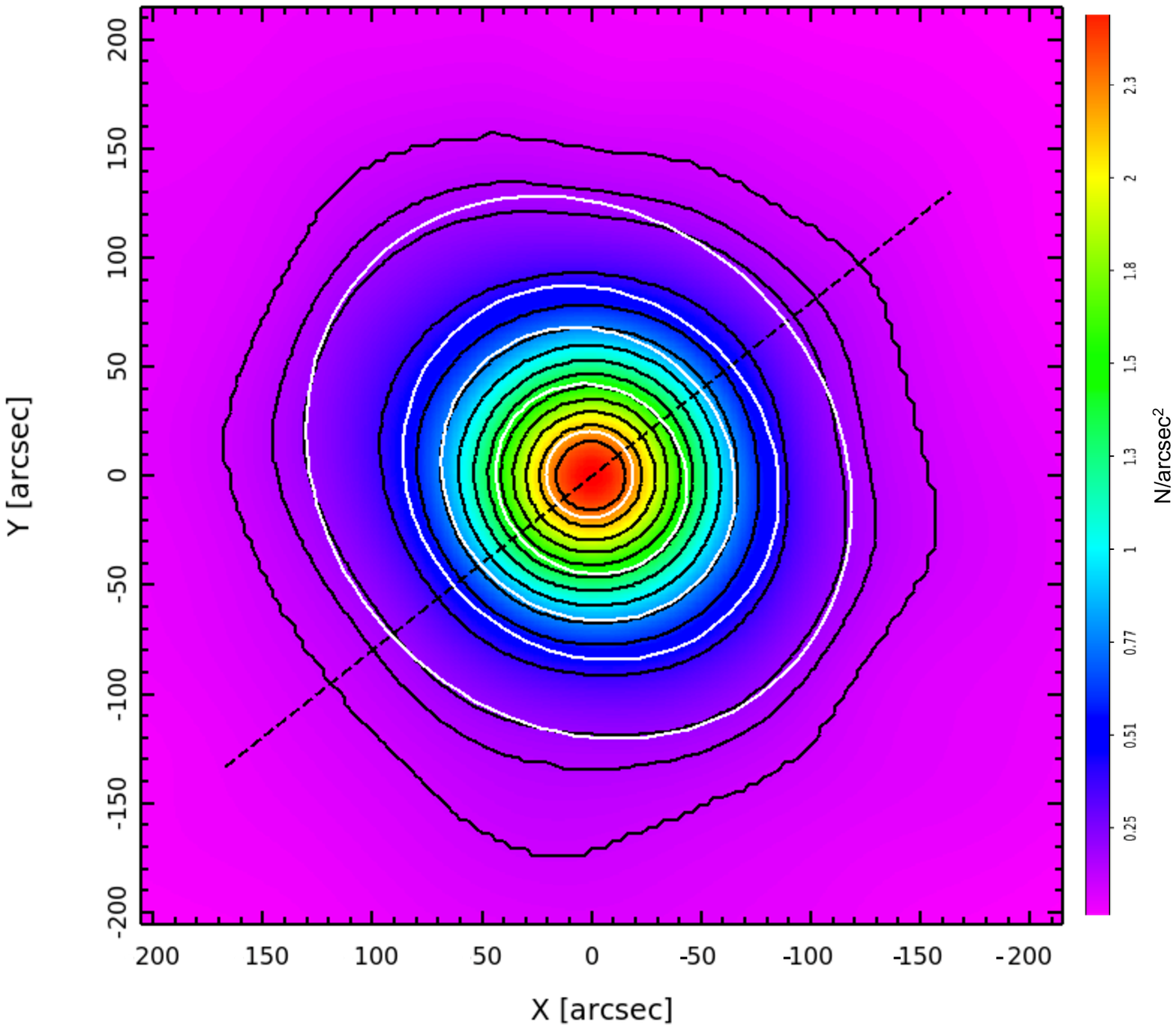}
\caption{Stellar density map (number of stars per square arcsecond:
  see the color-bar) of the inner $200\arcsec\times 200\arcsec$ of M5,
  obtained from HST/ACS and ESO-WFI photometry. The solid black lines
  are isodensity contours, the white curves are their best-fit
  ellipses. The black dashed line marks the direction of the global
  rotation axis (with position angle of $145\arcdeg$).}
\label{fig_dens}
\end{figure}

\begin{figure}
\includegraphics[width=15cm]{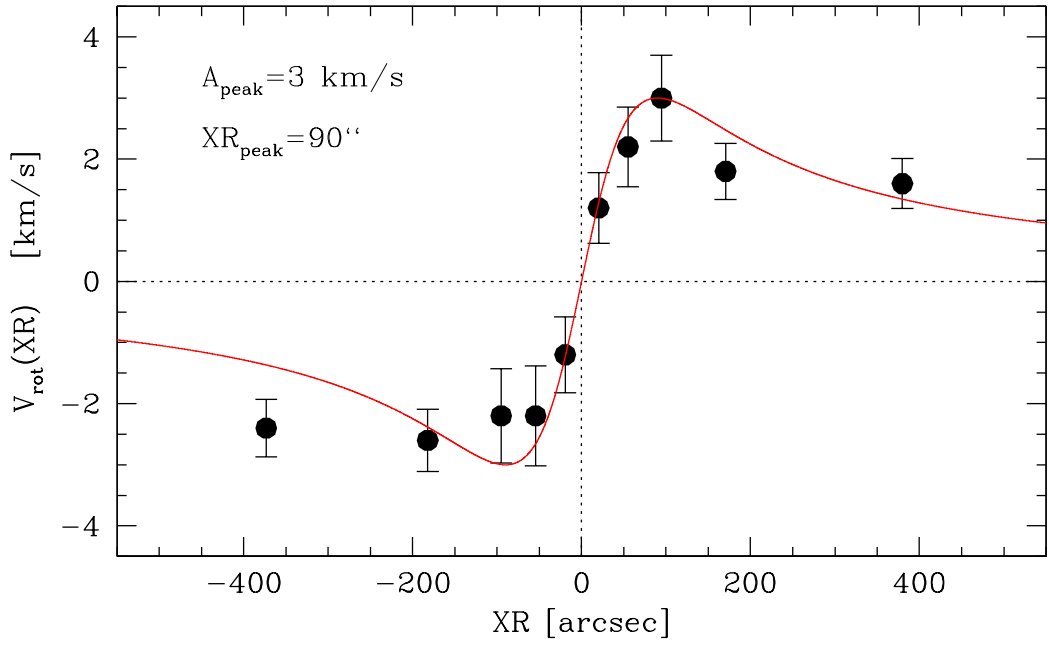}
\caption{Rotation curve of M5. The black circles mark the stellar mean
  velocity as a function of the projected distance on either side of
  the rotation axis (XR) for the intervals listed in Table
  \ref{tab_kin}. The red line, which well reproduces the observed
  curve, has the functional form expressed in equation
  (\ref{eq_curve}), with $A_{\rm peak}=3$ km s$^{-1}$ and ${\rm
    XR_{peak}}=90\arcsec$. }
\label{fig_rotcurve}
\end{figure}

\begin{figure}
\includegraphics[width=15cm]{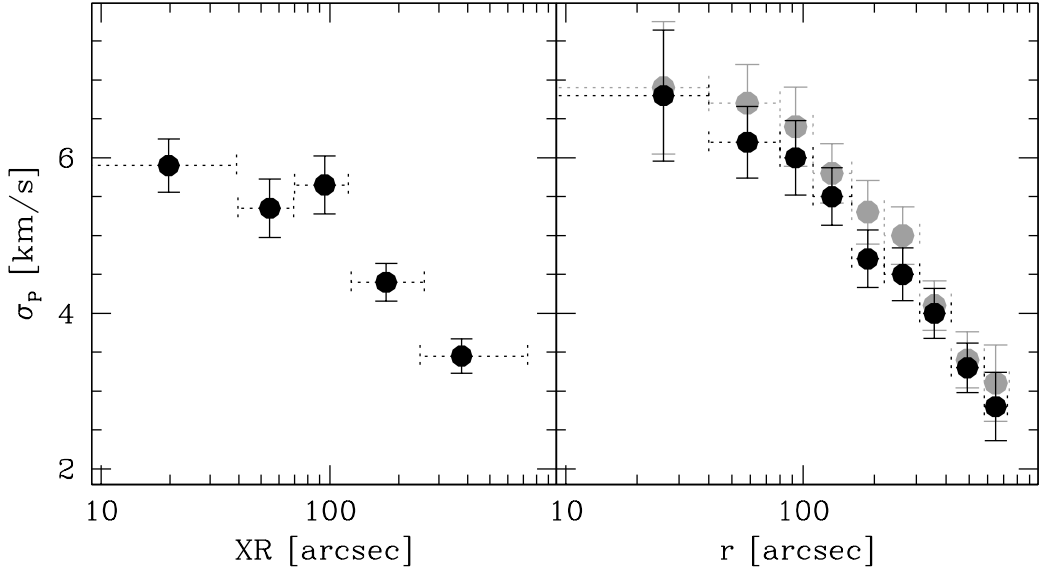}
\caption{Velocity dispersion profile of M5 obtained in two different
  projections.  \emph{Left-hand panel:} folded velocity dispersion
  profile determined in the same shells of projected distance from the
  rotation axis (XR) used for the rotation curve plotted in
  Fig. \ref{fig_rotcurve}. The corresponding values and error bars are
  listed in the last two columns of Table
  \ref{tab_kin}. \emph{Right-hand panel:} velocity dispersion profile
  obtained in concentric circular annuli around the cluster center
  (solid circles; see also Table \ref{tab_vdisp}). For sake of
  illustration, we also show the profile of the second velocity
  moment, which includes the effects of both rotation and velocity
  dispersion in each circular shell (grey circles and last two columns
  in Table \ref{tab_vdisp}).}
\label{fig_vdisp}
\end{figure}

With respect to previous works, our study has the advantage of being
based on a much larger statistics at $r>60\arcsec$. Hence, with the
exception of the central region, it provides the most solid and
precise determination of the rotation axis, rotation curve and
velocity dispersion profile of M5. Indeed, Fig. \ref{fig_pa} probably
shows the cleanest evidence so far of a constant value of PA$_0$ with
radius, testifying a coherent rotation and a reliable determination of
the central kinematics of this cluster.  The resulting rotation curve
is illustrated in Fig. \ref{fig_rotcurve}. This profile is well
reproduced by the analytic expression presented in equation
(\ref{eq_curve}), which is appropriate for cylindrical rotation and is
inspired by the structure of the velocity space of stellar systems
resulting from the process of violent relaxation \citep{lyndenbell67,
  gott73}. The observed peak rotation amplitude is $A_{\rm peak} \sim
3$ km s$^{-1}$ and is located at about $0.6 r_h$ from the center. The
radial distribution of the angular momentum is such that the behavior
in the central regions is consistent with solid-body rotation, while
in the outer portion of the radial range under consideration, it
declines smoothly.  Of course, kinematic information along the line of
sight provides exclusively a lower limit to the three-dimensional
rotation content, due to projection effects.

To study the relative importance of ordered versus random motions and
to quantify the role of rotation in shaping the geometry of a stellar
system, the ratio between the peak rotational velocity and the central
velocity dispersion is commonly used (for recent studies, see, e.g.,
\citealp{bianchini+13, kacharov+14, jeffreson+17}).  Since our data do
not sample the inner region of M5, we adopt the central velocity
dispersion $\sigma_0=7.3$ km s$^{-1}$ quoted by \citet{kamann+18},
finding $V_{\rm peak}/\sigma_0=0.4$. As discussed in
Sect. \ref{sec_dens}, we adopt $e=0.15$ for the cluster
ellipticity. In a plot of $V_{\rm peak}/\sigma_0$ versus the
ellipticity, M5 is the GC with largest rotational support that exactly
locates on the line of isotropic oblate rotators viewed edge-on (see,
e.g., Figure 14 in \citealp{bianchini+13}).  Hence, on the basis of
this simple argument, we suggest that the observed rotation amplitude
is likely close to the three-dimensional one (i.e., the stellar system
is observed on a line-of-sight which is close to the edge-on
projection), and the flattening of this cluster could be explained by
its own internal rotation.

\begin{figure}
\includegraphics[width=15cm]{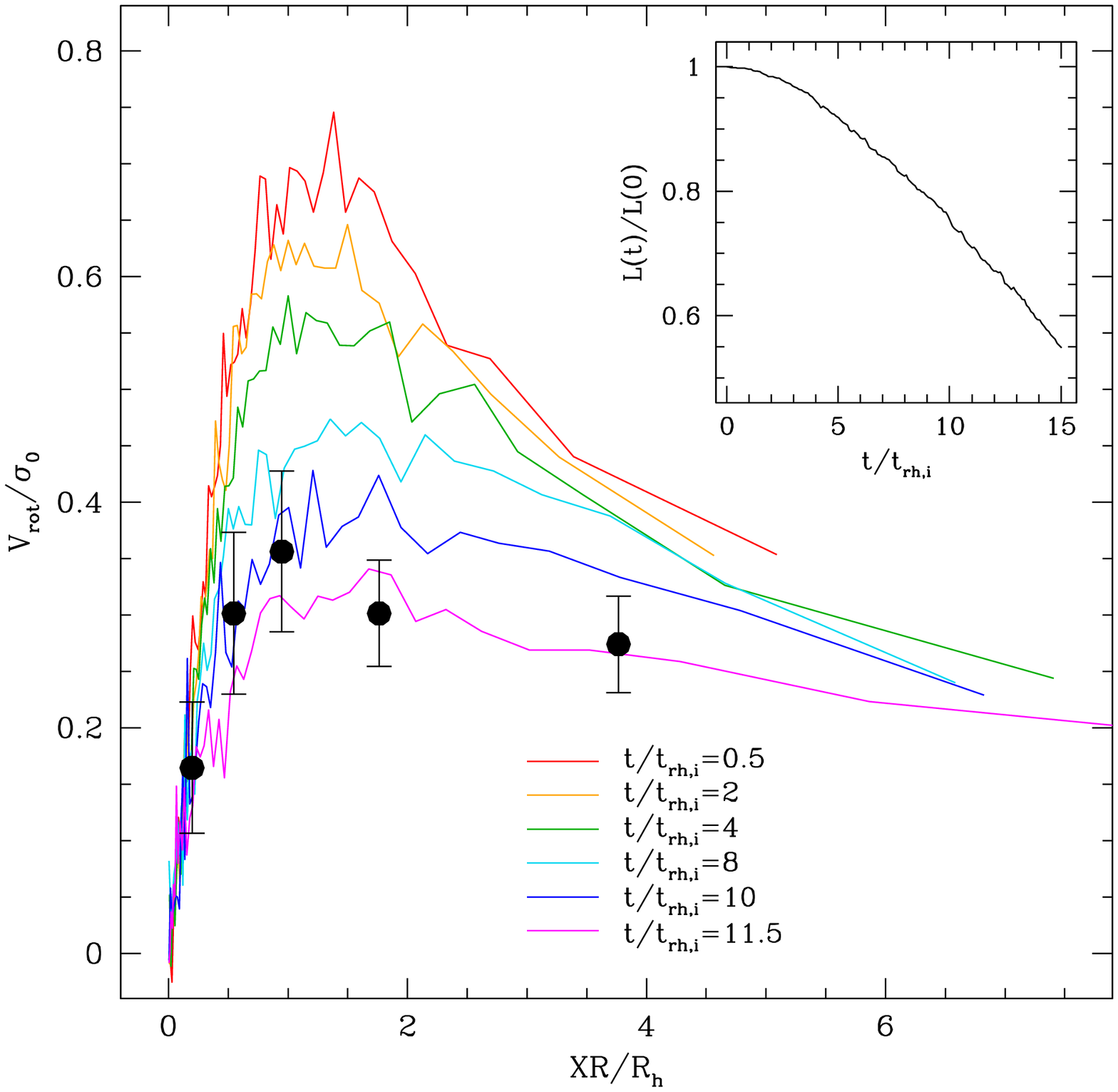}
\caption{Time evolution of the radial profile of $V_{\rm
    rot}/\sigma_0$ from one of the N-body simulations of \citet[][the
    VBrotF04 model, see their Table 1]{tiongco+16}. The radial
  distances from the rotation axis (XR) are express in units of the
  projected half-mass radius ($R_h=100\arcsec$; from
  \citealp{miocchi+13}).  An inclination angle between the
  line-of-sight and the rotation axis of $20\arcdeg$ is assumed in the
  calculation of the radial profiles. Each line shows the radial
  profile of $V_{\rm rot}/\sigma_0$ calculated at different times (see
  labels, where $t_{\rm rh,i}$ is the cluster's initial half-mass
  relaxation time). Each profile is calculated by combining three
  snapshots around the desired time. The black filled circles show the
  observed radial profile of M5: each point is the average of the
  $V_{\rm rot}$ values determined on the two sides of the rotation
  axis (see Table \ref{tab_kin}), normalized to the central velocity
  dispersion from \citet{kamann+18}. The inset shows the time
  evolution of the cluster's total angular momentum (L), normalized to
  its initial value.}
\label{fig_simu}
\end{figure}

In a forthcoming article, we will present a complete investigation
based on a global, self-consistent, axisymmetric dynamical model,
characterized by differential rotation and anisotropy in the velocity
space \citep[e.g.][]{varri+12}, coupled with appropriate N-body
simulations \citep[e.g.][]{tiongco+16, tiongco+17, tiongco+18}.
Nonetheless, here we present a first comparison between the radial
profile of the ratio $V_{\rm rot}/\sigma_0$ and the time evolution of
such a kinematic observable, as resulting from a representative N-body
model from the survey recently conducted by \citet{tiongco+16,
  tiongco+18}. Such a comparison, which is illustrated in Figure
\ref{fig_simu}, supports the conclusion that M5 has already
experienced the effects of two-body relaxation and angular momentum
transport over the course of several initial half-mass relaxation
times ($t_{\rm rh,i}$).  This simple analysis should be intended only
as a proof-of-concept that, in most cases, the angular momentum
measured in present-day GCs represent a lower limit of the amount they
possessed at birth (for the time evolution of the total angular
momentum of the model, see the figure inset).  We wish to emphasize
that this comparison did not require any ad-hoc tailoring of the
initial conditions of the N-body model, and involved exclusively a
simple exploration of the projected observables over different
lines-of-sight. The inclination angle adopted in the figure
($20\arcdeg$) is in qualitative agreement with the conclusion of an
nearly edge-on view of the system discussed above. However, this value
should be considered only as a representative example of a range of
acceptable values, while a definitive assessment requires a full
investigation of the degeneracy between intrinsic rotation and
projection effects, which will be presented in the forthcoming
dynamical study.

Rotation patterns as clear as those found in M5 have been detected
just in a few other cases so far (see the cases of NGC 4372 in
\citealt{kacharov+14}, and 47 Tucanae in
\citealt{bellini+17}). However, evidence of systemic rotation
signatures is mounting, with the most recent results for 9 GCs
presented in \citet[][but see also \citealt{bellazzini+12,
    fabricius+14, kacharov+14, kimmig+15, bellini+17, boberg+17,
    kamann+18}, and references therein]{ferraro+18a}.  This suggests
that possibly most (if not all) Galactic GCs are characterized by some
degree of internal rotation, that migh be the residual of a much
larger amount of ordered motions imprinted at birth and gradually
dissipated via angular momentum transport (due to two-body relaxation)
and mass loss \citep[e.g.,][see also Fig. \ref{fig_simu}]{fiestas+06,
  tiongco+17}.  Then, once combined with independent measures of the
level of dynamical evolution determined, e.g., from the radial
distribution of blue straggler stars (see \citealp{ferraro+09_m30,
  ferraro+12,ferraro+18b, lanzoni+16, raso+2017}), these signals may
be used to clarify the formation and evolutionary histories of GCs,
and the relative role of rotation.  This outlook substantiates the
urgency of multi-spectrograph studies of Galactic GCs (as the MIKiS
Survey), sensible enough to detect even weak rotation signals in these
systems. It also motivates the investment of renewed energies in the
theoretical investigation of the role of angular momentum in
collisional stellar dynamics, with appropriate equilibrium and
evolutionary dynamical models.
    
\acknowledgements{We thank the anonymous referee for useful comments that helped improving the presentation of the paper. F.R.F. acknowledges the ESO Visitor Programme for
  the support and the warm hospitality at the ESO Headquarter in
  Garching (Germany) during the period when part of this work was
  performed. ALV acknowledges support from a Marie Sklodowska-Curie Fellowship (MSCA-IF-EFRI 658088).}

\newpage
\begin{table}[h!]
\begin{center}
\begin{tabular}{lll}
\hline
Program ID  & Grating & PI\\                    
\hline
 193.D-0232 & HR21 & (Ferraro)     \\
 073.D-0695 &  HR5 & (Recio Blanco)\\ 
 088.B-0403 &  HR9 & (Lucatello)   \\
 073.D-0211 & HR11 & (Carretta)    \\ 
 087.D-0230 & HR12 & (Gratton)     \\ 
 073.D-0211 & HR13 & (Carretta)    \\ 
 087.D-0276 & HR15 & (D'Orazi)      \\
\hline
\hline
\end{tabular}
\end{center}
\caption{Summary of the FLAMES datasets used to derive the internal
  kinematics of M5.  The MIKiS Survey sample corresponds to Program ID
  193.D-0232, while the others have been retrieved from the ESO
  archive.}
\label{tab_datasets}
\end{table}

~

~

~

~

\begin{table}[h!]
\begin{center}
\begin{tabular}{rrrrccccc}
\hline
$r_i$  & $r_e$ & $r_m$ &  $N$ & PA$_0$ & $A_{\rm rot}$ & $P_{\rm KS}$ & $P_{\rm Stud}$ & n-$\sigma_{\rm ML}$\\
\hline
   20  &   40  &  29.4 &   89 &  163  &   2.3 & $7.5\times 10^{-2}$ &  $>90.0$  & 1.4  \\ 
   40  &   70  &  54.0 &  105 &  145  &   2.0 & $1.0\times 10^{-5}$ &  $>99.8$  & 4.2  \\ 
   70  &  110  &  88.8 &  118 &  144  &   2.3 & $1.1\times 10^{-2}$ &  $>99.8$  & 3.4  \\
  110  &  150  & 128.3 &  108 &  148  &   2.5 & $5.0\times 10^{-5}$ &  $>99.8$  & 3.9  \\
  150  &  220  & 182.0 &  111 &  151  &   1.9 & $1.5\times 10^{-4}$ &  $>99.8$  & 4.8  \\
  220  &  320  & 268.0 &  107 &  144  &   2.3 & $1.0\times 10^{-5}$ &  $>99.8$  & 4.7  \\
  320  &  600  & 426.2 &  141 &  145  &   1.4 & $3.1\times 10^{-3}$ &  $>99.8$  & 3.6  \\
\hline
\end{tabular}
\end{center}
\caption{Rotation signatures detected in circular annuli around the
  cluster center. For each annulus the table lists: the inner and
  outer radius ($r_i$ and $r_e$) in arcseconds, the mean radius and
  the number of stars in the bin ($r_m$ and $N$, respectively), the
  position angle of the rotation axis (PA$_0$), the rotation amplitude
  (A$_{\rm rot}$), the Kolmogorov-Smirnov probability that the two
  samples on each side of the rotation axis are drawn from the same
  parent distribution ($P_{\rm KS}$), the t-Student probability that
  the two RV samples have different means ($P_{\rm Stud}$), and the
  significance level (in units of n-$\sigma$) that the two means are
  different following a Maximum-Likelihood approach (n-$\sigma_{\rm
    ML}$).}
\label{tab_vrot_anelli}
\end{table}

~

~

~

~

\begin{table}[h!]
\begin{center}
\begin{tabular}{rrrrccccc}
\hline
$r_i$  & $r_e$ & $r_m$ &  $N$ & PA$_0$ & $A_{\rm rot}$ & $P_{\rm KS}$ & $P_{\rm Stud}$ & n-$\sigma_{\rm ML}$\\
\hline
  6  &   727  & 186.3 &  823 &  145  &    2.0       & $1.0 \times 10^{-17}$ &  $>99.8$  & 9.9  \\ 
\hline
\end{tabular}
\end{center}
\caption{The same as in Table \ref{tab_vrot_anelli}, but for the
  entire radial range ($6\arcsec-727\arcsec$) covered by the
  observations and after fixing the position angle of the rotation
  axis to PA$_0=145\arcdeg$.}
\label{tab_vrot_glob_PA145}
\end{table}

~

~

~

~

\begin{table}[h!]
\begin{center}
\begin{tabular}{rrrrccrrcccc}
\hline
XR$_i$ & XR$_e$ & XR$_m+$ &  $N+$ & $V_{\rm rot}+$ & $\epsilon_{V+}$ & XR$_m-$ &  $N-$ & $V_{\rm rot}-$ & $\epsilon_{V-}$ & $\sigma_P$(XR) & $\epsilon_{\sigma_P}$ \\
\hline
   0 &  40 &  20.2 & 131  & 1.2 & 0.6 &  -19.3 & 134 & -1.2 & 0.6 & 5.9 & 0.3 \\
  40 &  70 &  55.0 &  69  & 2.2 & 0.6 &  -54.3 &  55 & -2.2 & 0.8 & 5.4 & 0.4 \\
  70 & 120 &  94.6 &  64  & 3.0 & 0.7 &  -95.2 &  65 & -2.2 & 0.8 & 5.6 & 0.4 \\
 120 & 250 & 170.6 &  91  & 1.8 & 0.5 & -182.1 &  79 & -2.6 & 0.5 & 4.4 & 0.2 \\
 250 & 727 & 379.9 &  65  & 1.6 & 0.4 & -373.3 &  65 & -2.4 & 0.5 & 3.4 & 0.2 \\
\hline
\end{tabular}
\end{center}
\caption{Rotation curve and folded velocity dispersion profile of M5
  in the rotated system (XR,YR). For five intervals of projected
  distances from the rotation axis (XR), the tables lists: the inner
  and outer absolute limits of each bin (XR$_i$ and XR$_e$) in
  arcseconds, the mean distance, number of stars, average velocity and
  its error (in km s$^{-1}$) on the positive side of the XR axis
  (columns 3--6), the same for the negative side of the XR axis
  (columns 7--10), and the folded velocity dispersion and its error
  (columns 11-12).}
\label{tab_kin}
\end{table}

\begin{table}[h!]
\begin{center}
\begin{tabular}{rrrrcccc}
\hline
$r_i$ & $r_e$ & $r_m$ & $N$ & $\sigma_P(r)$ &  $\epsilon_{\sigma}$ & $\tilde\sigma_P(r)$ &  $\epsilon_{\tilde\sigma}$ \\
\hline
   6 &  40 &  25.8 & 116 & 6.8 & 0.8 & 6.9 & 0.9 \\  
  40 &  80 &  58.2 & 132 & 6.2 & 0.5 & 6.7 & 0.5 \\   
  80 & 110 &  93.0 &  91 & 6.0 & 0.5 & 6.4 & 0.5 \\ 
 110 & 160 & 132.2 & 126 & 5.5 & 0.4 & 5.8 & 0.4 \\ 
 160 & 220 & 187.2 &  93 & 4.7 & 0.4 & 5.3 & 0.4 \\ 
 220 & 310 & 262.5 &  96 & 4.5 & 0.3 & 5.0 & 0.4 \\ 
 310 & 420 & 357.6 &  88 & 4.0 & 0.3 & 4.1 & 0.3 \\ 
 420 & 580 & 492.5 &  58 & 3.3 & 0.3 & 3.4 & 0.4 \\ 
 580 & 727 & 648.3 &  23 & 2.8 & 0.4 & 3.1 & 0.5 \\ 
\hline
\end{tabular}
\end{center}
\caption{Velocity dispersion and second velocity moment profiles of M5
  determined in circular annuli around the cluster center: internal
  and external radius of each annulus in arcseconds (columns 1,2),
  average cluster-centric distance and number of stars in the bin
  (columns 3,4), projected velocity dispersion (km s$^{-1}$) and its
  uncertainty (columns 5,6), projected second velocity moment and its
  uncertainty (columns 7,8).}
\label{tab_vdisp}
\end{table}

\newpage 

\end{document}